\documentstyle[12pt,psfig]{article}
\makeatletter
\def\@cite#1#2{{[{#1}]\if@tempswa\typeout
{IJCGA warning: optional citation argument
ignored: `#2'} \fi}}

\newcount\@tempcntc
\def\@citex[#1]#2{\if@filesw\immediate\write\@auxout{\string\citation{#2}}\fi
  \@tempcnta\z@\@tempcntb\m@ne\def\@citea{}\@cite{\@for\@citeb:=#2\do
    {\@ifundefined
       {b@\@citeb}{\@citeo\@tempcntb\m@ne\@citea\def\@citea{,}{\bf ?}\@warning
       {Citation `\@citeb' on page \thepage \space undefined}}%
    {\setbox\z@\hbox{\global\@tempcntc0\csname b@\@citeb\endcsname\relax}%
     \ifnum\@tempcntc=\z@ \@citeo\@tempcntb\m@ne
       \@citea\def\@citea{,}\hbox{\csname b@\@citeb\endcsname}%
     \else
      \advance\@tempcntb\@ne
      \ifnum\@tempcntb=\@tempcntc
      \else\advance\@tempcntb\m@ne\@citeo
      \@tempcnta\@tempcntc\@tempcntb\@tempcntc\fi\fi}}\@citeo}{#1}}
\def\@citeo{\ifnum\@tempcnta>\@tempcntb\else\@citea\def\@citea{,}%
  \ifnum\@tempcnta=\@tempcntb\the\@tempcnta\else
   {\advance\@tempcnta\@ne\ifnum\@tempcnta=\@tempcntb \else
\def\@citea{--}\fi
    \advance\@tempcnta\m@ne\the\@tempcnta\@citea\the\@tempcntb}\fi\fi}
\makeatother
\newenvironment{Eqnarray}%
     {\arraycolsep 0.14em\begin{eqnarray}}{\end{eqnarray}}

\def\simgt{\stackrel{>}{{}_\sim}}
\def\be{\begin{equation}}
\def\ee{\end{equation}}
\def\bear{\be\begin{array}}
\def\eear{\end{array}\ee}
\def\bea{\begin{Eqnarray}}
\def\eea{\end{Eqnarray}}

\def\lsim{\mathrel{\raise.3ex\hbox{$<$\kern-.75em\lower1ex\hbox{$\sim$}}}}
\def\gsim{\mathrel{\raise.3ex\hbox{$>$\kern-.75em\lower1ex\hbox{$\sim$}}}}
\def\ifmath#1{\relax\ifmmode #1\else $#1$\fi}
\def\ls#1{\ifmath{_{\lower1.5pt\hbox{$\scriptstyle #1$}}}}

\def\beq{\begin{equation}}
\def\eeq{\end{equation}}
\def\beqa{\begin{Eqnarray}}
\def\eeqa{\end{Eqnarray}}

  \def\simgt{\stackrel{>}{{}_\sim}}
\def\bma#1{\mbox{\boldmath{$#1$}}}

\def\baselinestretch{1}
\begin{document}
\def\IJMPA #1 #2 #3 {{\sl Int.~J.~Mod.~Phys.}~{\bf A#1}\ (19#2) #3$\,$}
\def\MPLA #1 #2 #3 {{\sl Mod.~Phys.~Lett.}~{\bf A#1}\ (19#2) #3$\,$}
\def\NPB #1 #2 #3 {{\sl Nucl.~Phys.}~{\bf B#1}\ (19#2) #3$\,$}
\def\PLB #1 #2 #3 {{\sl Phys.~Lett.}~{\bf B#1}\ (19#2) #3$\,$}
\def\PR #1 #2 #3 {{\sl Phys.~Rep.}~{\bf#1}\ (19#2) #3$\,$}
\def\JHEP #1 #2 #3 {{\sl JHEP}~{\bf #1}~(19#2)~#3$\,$}
\def\PRD #1 #2 #3 {{\sl Phys.~Rev.}~{\bf D#1}\ (19#2) #3$\,$}
\def\PTP #1 #2 #3 {{\sl Prog.~Theor.~Phys.}~{\bf #1}\ (19#2) #3$\,$}
\def\PRL #1 #2 #3 {{\sl Phys.~Rev.~Lett.}~{\bf#1}\ (19#2) #3$\,$}
\def\RMP #1 #2 #3 {{\sl Rev.~Mod.~Phys.}~{\bf#1}\ (19#2) #3$\,$}
\def\ZPC #1 #2 #3 {{\sl Z.~Phys.}~{\bf C#1}\ (19#2) #3$\,$}
\def\PPNP#1 #2 #3 {{\sl Prog. Part. Nucl. Phys. }{\bf #1} (#2) #3$\,$}

\catcode`@=11
\newtoks\@stequation
\def\subequations{\refstepcounter{equation}%
\edef\@savedequation{\the\c@equation}%
  \@stequation=\expandafter{\theequation}
  \edef\@savedtheequation{\the\@stequation}
  \edef\oldtheequation{\theequation}%
  \setcounter{equation}{0}%
  \def\theequation{\oldtheequation\alph{equation}}}
\def\endsubequations{\setcounter{equation}{\@savedequation}%
  \@stequation=\expandafter{\@savedtheequation}%
  \edef\theequation{\the\@stequation}\global\@ignoretrue

\noindent}
\catcode`@=12
\begin{titlepage}
\title{\bf 
Symmetry Nonrestoration at High Temperature in Little Higgs Models}
\author{ 
{\bf J.R. Espinosa$^{1}$\footnote{\baselineskip=16pt E-mail: {\tt
jose.espinosa@cern.ch}}},
{\bf M. Losada$^{2}$\footnote{\baselineskip=16pt E-mail: {\tt
malosada@uan.edu.co}}} and {\bf A. 
Riotto$^{3}$\footnote{\baselineskip=16pt 
E-mail: {\tt
antonio.riotto@pd.infn.it}}}\\ 
\hspace{3cm}\\
 $^{1}$~{\small IFT-UAM/CSIC, Fac. Ciencias UAM, Cantoblanco, 28049 
Madrid, Spain}
\hspace{0.3cm}\\
 $^{2}$~{\small Centro de Investigaciones, Univ. Antonio Nari\~no, 
Cll. 58A No 37-94, Bogot\'a, Colombia}
\hspace{0.3cm}\\
 $^{3}$~{\small INFN, Sezione di Padova, via Marzolo 8, Padova I-35131, 
Italy}} 
\maketitle 
\def\baselinestretch{1.15} 
\begin{abstract}
\noindent
A detailed study of the high temperature dynamics of the scalar 
sector of Little Higgs scenarios, proposed to stabilize the electroweak 
scale,
shows that the electroweak gauge symmetry
remains broken even at temperatures much larger than the
electroweak scale. Although we give explicit results for a
particular modification of the Littlest Higgs model, we expect that
the main features are generic.  As a spin-off, we introduce a novel
way of dealing with scalar fluctuations in nonlinear sigma models,
which might be of interest for phenomenological applications.
\end{abstract}

\thispagestyle{empty} 
\vskip2cm
\leftline{September 2004} \leftline{}

\vskip-20cm \rightline{} 
\rightline{IFT-UAM/CSIC-04-44} 
\rightline{CI-UAN/04-07FT}
\rightline{hep-ph/0409070} \vskip3in

\end{titlepage}
\setcounter{footnote}{0} \setcounter{page}{1}
\newpage
\baselineskip=20pt

\section{Introduction}

Little Higgs (LH) models \cite{AH,Littlest,Schmaltz,Wacker} provide a new
scenario for electroweak symmetry breaking which stabilizes the
electroweak scale. It is well known that symmetries in the Standard Model
(SM) Lagrangian force gauge bosons and fermions to be massless. The main
feature of LH models is the inclusion of additional global symmetries
which force the Higgs boson mass to be zero. The incorporation of explicit
violations of these symmetries in a precise way explains why the ratio of
the Higgs mass to the cutoff of the theory, $m_H^2/\Lambda^2$, is small:
the Little Higgs boson is a naturally light pseudo-Goldstone boson
\cite{Littlest}.

Below the cutoff scale $\Lambda\sim 10$ TeV, there are additional
particles (with masses of order $f\sim 1$ TeV) which cancel the
quadratically divergent contributions to the Higgs boson mass from SM
particles (or at least the most dangerous ones). Beneath the TeV scale the
effective degrees of freedom are those of the Standard Model. The initial
implementation of this idea was based on a SU(5)/SO(5) nonlinear sigma
model which contained a gauged $[SU(2)\times U(1)]^2$ subgroup. However,
this model is in conflict with low-energy precision electroweak
measurements and from direct searches for a $Z'$ boson 
\cite{Han,Csaki,Hewett,Burdman,muchun}, both problems being related to the 
additional $U(1)$ gauge group. Although
this problem is not of direct concern for our purposes, we focus in this
paper  on a Little Higgs model which is a simple variation of the 
so-called Littlest
one, having only a gauged $SU(2)\times SU(2)\times U(1)$ group
\cite{Peskin}. The performance of the model with respect to 
electroweak fits is improved while the smallness of the $g'$ coupling
tames the remnant quadratic divergence in the Higgs boson mass associated
to the $U(1)_Y$ gauge group.\footnote{An interesting alternative has been 
proposed recently \cite{ChengLow} in
which a $T$-parity is imposed on the nonlinear sigma model such that
it eliminates the strongest constraints from tree-level processes to
electroweak observables. In any case, most of our findings
are expected to be generic in Little Higgs models and should hold also 
for models with $T$-parity.}

In this paper we study the finite temperature behaviour of Little Higgs
models. The main motivation is related to the peculiar way in which such
models cancel the quadratically divergent corrections to the Higgs boson
mass. As is well known, in the early Universe the Higgs scalar field gets
an effective thermal mass (squared)  $m_{eff}^2\sim T^2$ from interactions
with the ambient hot plasma. It is this positive $m_{eff}^2$ which is
responsible for symmetry restoration at high temperature \cite{symrest}.  
This effective mass can be computed diagrammatically using well
established finite temperature techniques;  the
self-energy diagrams which  are quadratically divergent at $T=0$ are 
precisely the ones
that give a contribution to the Higgs thermal mass. The correspondence is
very simple \cite{coes}: a one-loop bosonic self-energy diagram that gives
$\delta m_h^2= \kappa \Lambda^2/(16\pi^2)$, where $\Lambda$ is an UV
cutoff, produces $\delta m_{eff}^2= \kappa T^2/12$ at finite temperature.
For a fermionic one-loop self-energy diagram, if the zero $T$ result is
$\delta m_h^2= -\kappa \Lambda^2/(16\pi^2)$, the finite $T$ contribution
of the same diagram will be $\delta m_{eff}^2= \kappa T^2/24$. For
instance, in the Standard Model one gets the well known quadratically
divergent correction to the Higgs mass
\be
\delta m_h^2 ={3 \Lambda^2\over 16 \pi^2}\left[
2\lambda + {1\over 2} g^2 + {1\over 4}(g^2+{g'}^2)-2 h_t^2
\right]\ ,
\label{veltman}
\ee
where $\lambda$ is the Higgs quartic coupling (normalized so as to have 
$m_h^2=2\lambda v^2$ with $v=246$ GeV), $g$ is the $SU(2)$ gauge 
coupling, $g'$ the $U(1)_Y$ coupling and $h_t$ the top Yukawa coupling.
The result (\ref{veltman}) translates, at $T\neq 0$, into
\be
\delta m_{eff}^2 = {T^2\over 4}\left[
2\lambda + {1\over 2} g^2 + {1\over 4}(g^2+{g'}^2)+ h_t^2
\right]\ .
\ee
In the MSSM, although the sum of quadratically divergent corrections to 
$m_h^2$ cancels between bosonic and fermionic degrees of freedom, the 
corresponding corrections at $T\neq 0$ do not  and 
one gets symmetry restoration also in that case.

The defining property of Little Higgs models is that quadratically
divergent corrections to $m_h^2$ cancel sector by sector in the model,
between particles {\it of the same statistics}. At $T\neq 0$ this leads
one to expect that $\delta m_{eff}^2$ will also be zero in this type of
models. Notice though that this will only happen for sufficiently large
temperatures, such that the heavy partners of the SM particles are
thermally produced and populate the hot plasma. At low temperatures we
expect the heavy particles with mass $\sim f$ to be Boltzmann decoupled,
and we find the same SM $T^2$ dependence on the Higgs mass, with the
electroweak symmetry being restored as usual. (That is, the electroweak
phase transition will take place as in the SM, at $T\sim 100$ GeV.) As the
temperature approaches $f\sim 1$ TeV the new particles introduced in
Little Higgs models will be in thermal equilibrium in the plasma, the
thermal Higgs mass will drop to zero and the electroweak symmetry will be
broken again\footnote{It is interesting to confront this result with the 
negative expectations of ref.~\cite{nogo}, which claimed that symmetry 
nonrestoration could only be obtained at the price of hierarchy 
problems, while we find symmetry nonrestoration precisely due to the good 
UV properties of Little Higgs models. This contradiction is resolved by 
noting that the results of \cite{nogo} do not apply in nonlinear sigma 
models.}. Although these expectations will prove to be correct, they
concern the behaviour of the Higgs potential, $V(h)$, for small values of
the Higgs field $h$. In order to study what happens to the minimum of 
the Higgs potential at $T\simeq f\sim 1$
TeV, we will need to study larger values of $h$. 

Without knowing the UV theory that supersedes the Little Higgs model
beyond $\Lambda\simeq 4\pi f$ we cannot study the behaviour of our theory
at $T\geq \Lambda/\pi\sim 4 f$ (at such temperatures the free-energy
contribution from particles with mass $\sim \Lambda$ is no longer
negligible). With this limitation in mind, we would like to explore in
this paper the behaviour of the Higgs potential of Little Higgs models at
finite temperature.  We will confirm the expected behaviour described in
the previous paragraph and discover some peculiar features in the
temperature evolution of $V(h)$. We present detailed results for the model
of ref.~\cite{Peskin} but expect that the main features we find are
generic as they are based on the defining properties of Little Higgs
models.

In section~2 we present the model and our notation. Section~3 describes in
detail the structure of the zero temperature effective potential at
 one-loop order,
paying particular attention to the treatment of the contributions from
scalar fluctuations. Section~4 then goes on to compute the finite
temperature effective potential and describes some interesting features of
its temperature evolution. In particular, we 
discover that the electroweak gauge symmetry remains broken even if the
system is heated up to temperatures larger than the electroweak scale.
In section~5 we
present some conclusions. Appendix~A contains explicit expressions for the
mass matrices of the different species of particles in the model,
calculated to all orders in the Higgs background, which are necessary to
compute the one-loop effective potential. Appendix~B deals with the
behaviour of the potential along the direction of the triplet field
contained in the model.

\section{The Little Higgs Model}

The model is based on the $SU(5)/SO(5)$ nonlinear sigma model of
\cite{Littlest} (the Littlest Higgs), modified according to \cite{Peskin}.  
The spontaneous breaking of $SU(5)$ down to $SO(5)$ is produced by the
vacuum expectation value (vev) of a $5\times 5$ symmetric matrix $\Phi$
[which transforms under $SU(5)$ as $\Phi\rightarrow U\Phi U^T$] for
instance when $\langle\Phi\rangle=I_5$ (we denote by $I_n$ the $n\times n$
identity matrix). This breaking of the global $SU(5)$ symmetry produces 14
Goldstone bosons among which lives the scalar Higgs field. Instead of
working on the background $\langle\Phi\rangle=I_5$ we follow
\cite{Littlest} and, making a basis change, we choose
$\langle\Phi\rangle=\Sigma_0$ where
\beq
\Sigma_0 =\left(\begin{array}{ccc}
0 &0&I_2\\
0&1&0\\
I_2&0&0
\end{array}\right)\ .
\label{vev}
\eeq
Calling $U_0$ the $SU(5)$ matrix that performs this change of basis, we
have $\Sigma_0=U_0U_0^T$ while all the group generators change as
$t_a=U_0t_a^{(0)}U_0^\dagger$ [$t_a^{(0)}$ are the generators in the
original basis]. The unbroken $SO(5)$ generators satisfied the obvious
relation $T_a^{(0)}+T_a^{(0)T}=0$ and multiplying on the left by $U_0$ and
on the right by $U_0^T$ we arrive at the condition
\be
T_a\Sigma_0+\Sigma_0T_a^T=0\ ,
\ee
for the generators in the new basis (a condition which is immediate to
obtain alternatively just by requiring invariance of $\Sigma_0$). In the
original basis the broken generators obviously satisfy
$X_a^{(0)}=X_a^{(0)T}$.  Multiplying again by $U_0$ and $U_0^T$ one gets 
in the transformed basis
\be
X_a\Sigma_0=\Sigma_0 X_a^T\ .
\ee
The Goldstone bosons 
can be parametrized through the nonlinear sigma model field
\beq
\Sigma = e^{i \Pi/f} \Sigma_0 e^{i \Pi^T/f} = e ^{2i\Pi/f}\Sigma_0,
\label{Sigma}
\eeq
where $\Pi = \sum_a \Pi^a  X^a$. The model assumes a gauged  
$SU(2)_1\times SU(2)_2\times U(1)_Y$ subgroup 
of $SU(5)$ with generators 
\beq
Q_1^a = \left(\begin{array}{cc}
\sigma^a/2 &0_{2\times 3}\\
0_{3\times 2}&0_{3\times 3}
\end{array}\right)\ , \hspace{0.5cm}  
Q_2^a = \left(\begin{array}{cc}
0_{3\times 3} &0_{3\times 2}\\
0_{2\times 3}&-\sigma^{a^{*}}/2
\end{array}\right)\ ,
\eeq
(where $\sigma^a$ are the Pauli matrices) and $Y={\mathrm
diag}(1,1,0,-1,-1)/2$. The vacuum expectation value in eq.~(\ref{vev})
additionally breaks $SU(2)_1\times SU(2)_2$ down to the SM $SU(2)$ group.  

The hermitian matrix $\Pi$ in eq.~(\ref{Sigma}) contains the Goldstone and 
(pseudo)-Goldstone bosons:
\beq
\Pi = \left(\begin{array}{ccc}
\xi &\frac{H^{\dagger}}{\sqrt 2}&\phi^{\dagger}\\
\frac{H}{\sqrt 2}&0&\frac{H^{*}}{\sqrt 2}\\
\phi&\frac{H^{T}}{\sqrt 2}&\xi^T
\end{array}\right)+{1\over \sqrt{20}}\zeta^0{\mathrm diag}(1,1,-4,1,1)\ ,
\label{pi}
\eeq
where $H=(h^0,h^+)$ is the Higgs doublet; $\phi$ is a complex $SU(2)$ 
triplet 
given by the symmetric $2\times 2$ matrix:
\be
\phi=\left[
\begin{array}{cc}
\phi^0 & {1\over \sqrt{2}} \phi^+\\
{1\over \sqrt{2}} \phi^+ & \phi^{++}
\end{array}
\right]\ ,
\ee
the field $\zeta^0$ is a singlet and finally, $\xi$ is the real triplet 
of 
Goldstone bosons associated to $SU(2)_1\times SU(2)_2\rightarrow SU(2)$ 
breaking:
\be
\xi={1\over 2} \sigma^a\xi^a=\left[
\begin{array}{cc}
{1\over 2}\xi^0 & {1\over \sqrt{2}} \xi^+\\
{1\over \sqrt{2}} \xi^- & -{1\over 2}\xi^{0}
\end{array}
\right]\ .
\ee

The kinetic part of the Lagrangian  is
\beq
{\cal L}_k = \frac{f^2}{8} 
{\mathrm Tr}[(D_{\mu}\Sigma)(D^{\mu}\Sigma)^{\dagger}]
\label{Lk}
\eeq
where
\beq
D_{\mu}\Sigma = \partial_{\mu} \Sigma - i\sum_{j=1}^2 g_j W_j^a(Q_j^a \Sigma + 
\Sigma Q_j^{a T})  - i g' B_Y(Y \Sigma + \Sigma Y^T).
\eeq

In this model additional fermions are introduced as a vector-like coloured
pair $\tilde{t}, \tilde{t}^c$ to cancel the quadratic divergence from top
loops (we neglect the other Yukawa couplings). The relevant part of the
Lagrangian containing the top Yukawa coupling is given by
\beq
{\cal L}_f = {1\over 2}\lambda_{1} f \epsilon_{ijk} \epsilon_{xy} \chi_i 
\Sigma_{jx} 
\Sigma_{ky} 
u'^{c}_{3} + \lambda_{2} f \tilde{t} \tilde{t}^c + h.c.,
\eeq
where $\chi_i = (b, t,\tilde{t})$, indices $i,j,k$ run from 1 to 3 and 
$x,y$ from 4 to 5, and $\epsilon$ is the completely antisymmetric tensor.

As explained in \cite{Littlest} considering gauge and fermion loops, one
sees that the Lagrangian should also include gauge invariant terms of the
form,
\bea
-{\cal L}_s =  V  & = & 2 c f^4 g_i^2 \sum_a 
{\mathrm Tr}[(Q_i^{a}\Sigma)(Q_i^{a}\Sigma)^*]  + 2 c f^4 {g'}^2 
{\mathrm Tr}[(Y\Sigma)(Y\Sigma)^*]\nonumber\\
& +& 4 c' f^4 \lambda_1^2 \epsilon^{wx}\epsilon_{yz}\Sigma_{iw}\Sigma_{jx}
\Sigma^{iy *}\Sigma^{jz *}\ ,
\label{potential}
\eea
with $c$ and $c'$ assumed to be constants of ${\cal O}(1)$.

This Lagrangian produces a mass of order $f$ for the gauge bosons
($W'$) associated to the broken (axial) $SU(2)$, for a vector-like
combination of $\tilde{t}$ and $u_3'$ and for the complex scalar $\phi$.  
The singlet $\zeta^0$ is a pure Goldstone [associated to the breaking
of the $U(1)$ symmetry left ungauged] that will play no significant role
in the discussion (it can be given a small mass to avoid phenomenological
problems by adding explicit breaking terms).  Finally, the Higgs boson
gets a small tree level mass of order ${g'}^2$ and a quartic coupling (not
suppressed by $g'$).

\section{Effective Potential (${\bma T} {\bma =} {\bma 0}$)}
\begin{figure}[t]
\vspace{1.cm} \centerline{
\psfig{figure=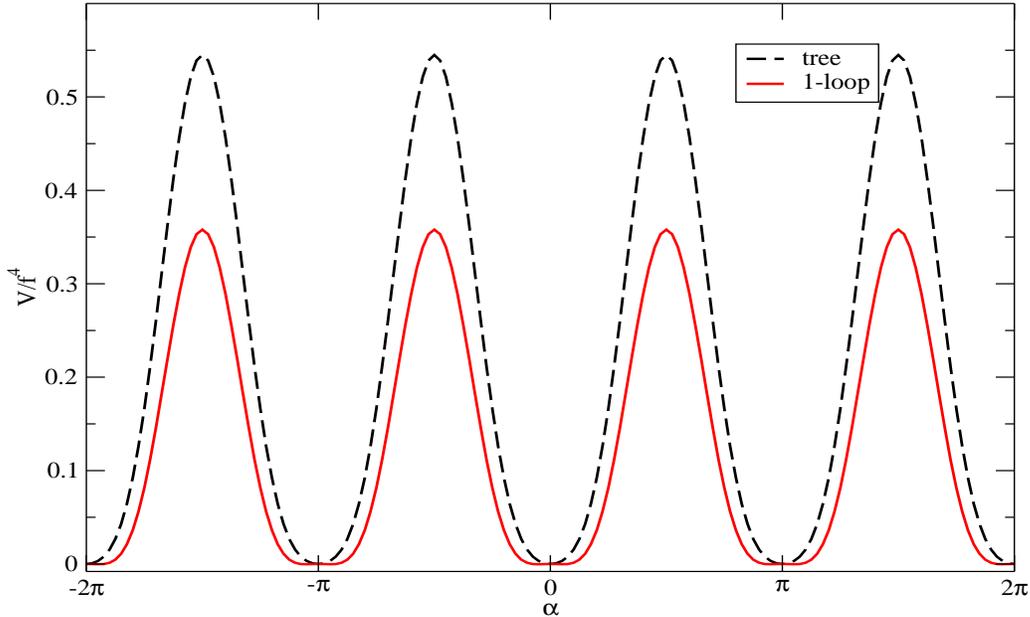,angle=-90,height=8cm,width=8cm,bbllx=4.cm,%
bblly=7.cm,bburx=20.cm,bbury=21.cm}}
\caption{\footnotesize
Periodic structure of the Little Higgs effective potential for $T=0$.}
\label{fig:periodV}
\end{figure}

From the Lagrangian (\ref{potential}) we can extract the tree-level
effective potential for the real scalar field $h\equiv \sqrt{2}{\mathrm
Re}(H)$.  In most previous papers an expansion in $h$ is performed. Here
we avoid doing this, except to illustrate a few aspects, and keep the
full dependence on $h$. The tree-level potential is a function of
\be
\langle\Sigma\rangle =\exp\left[2i\langle\Pi\rangle/f\right] \Sigma_0  
=\left[ I_5 + i \sqrt{2} \frac{\langle\Pi\rangle}{h} \sin 2 
\alpha- 4 \frac{\langle\Pi\rangle^2}{h^2}\sin^2 \alpha\right]\Sigma_0\ ,
\ee
where $\langle \Pi\rangle$ is only nonzero through $h$ and $\alpha\equiv 
h/(\sqrt{2}f)$. This is a periodic function 
and as a result the
potential
\be
V(h)={\mathrm Constant} + f^4\left[2c{g'}^2 s_\alpha^2 + {1\over 2}
\lambda_+ s_\alpha^4\right]\ ,
\ee
[where $\lambda_\pm\equiv c(g_1^2\pm g_2^2)\pm 16 c' \lambda_1^2$] is
invariant under $h\rightarrow h+n\pi 
f$. The minimum at the 
origin $h=0$
is replicated at $\pm n\pi f$ (see figure~\ref{fig:periodV}) with barriers
of height $(2c{g'}^2 +\lambda_+/2)f^4$ separating these minima. In each of 
them, in spite of appearances (the fact that $h\neq 0$) the electroweak
symmetry is unbroken. In fact one can show that the mass spectrum is the
same in all these vacua, in particular SM gauge bosons are massless. In 
other words, the order parameter for electroweak symmetry breaking is
$M_W$ rather than $\langle h\rangle$.\footnote{The periodicity extends in 
fact  to the plane $\{Re(h),Im(h)\}$ as the potential is a function of 
$|h|$ only. The minima at $n\pi f$ are then circles in that plane, 
which are nevertheless equivalent to the point at the origin $h=0$. 
One can think of $Re(h)$ and $Im(h)$ as orthogonal coordinates on a sphere 
to 
understand the periodic structure of the potential: different minima 
correspond to the same point on the sphere. }

Regarding the local properties of these minima, an expansion in 
powers of $h$ around $h=0$ gives:
\be
V(h)= {\mathrm Constant} + c{g'}^2 f^2h^2+{1\over 24}(3\lambda_+ - 4
c{g'}^2)h^4+...
\ee
As expected, the only\footnote{There 
are additional two-loop contributions
to the mass term \cite{Han}.}
mass term for $h$ is due to the $U(1)_Y$ gauge coupling $g'$. This offers 
an immediate possibility for electroweak symmetry breaking if one 
chooses $c<0$. In that case the minimum is at
\be
\langle s_\alpha^2 \rangle = {-2c{g'}^2\over \lambda_+}\ ,
\label{minimum}
\ee
or 
\be
\langle h^2\rangle\simeq {-12c{g'}^2 f^2\over (3\lambda_+-4c{g'}^2)}\ .
\ee

However, this tree-level breaking is problematic. The expansion of the 
potential to fourth order in $h$ and $t\equiv \sqrt{2}{\mathrm 
Re}(-i\phi)$ is:
\bea
V(h,t)&=&{\mathrm Constant} + c{g'}^2 f^2h^2+{1\over 24}(3\lambda_+ - 4
c{g'}^2)h^4+{1\over\sqrt{2}}\lambda_-fh^2t\nonumber\\
&+&(\lambda_++4c{g'}^2)f^2t^2-{1\over 6}(4\lambda_++17c{g'}^2)t^2h^2-
{2\over 3}(\lambda_++4c{g'}^2)t^4+...
\label{potentialht}
\eea
In the presence of the coupling $\lambda_-$, a vev for $h$ induces a tadpole 
for $t$ so that the previous minimum gets slightly displaced.
From (\ref{potentialht}) we get
\be
\langle t\rangle\simeq {-\lambda_-f\over2\sqrt{2}( 
\lambda_++4c{g'}^2)}{\langle h^2\rangle\over f^2}\ ,
\ee
but one cannot obtain $|\lambda_-|\ll \lambda_++4c{g'}^2$ and as a 
result $\langle t \rangle$ turns out to be too large.
\begin{figure}[t]
\vspace{1.cm} \centerline{
\psfig{figure=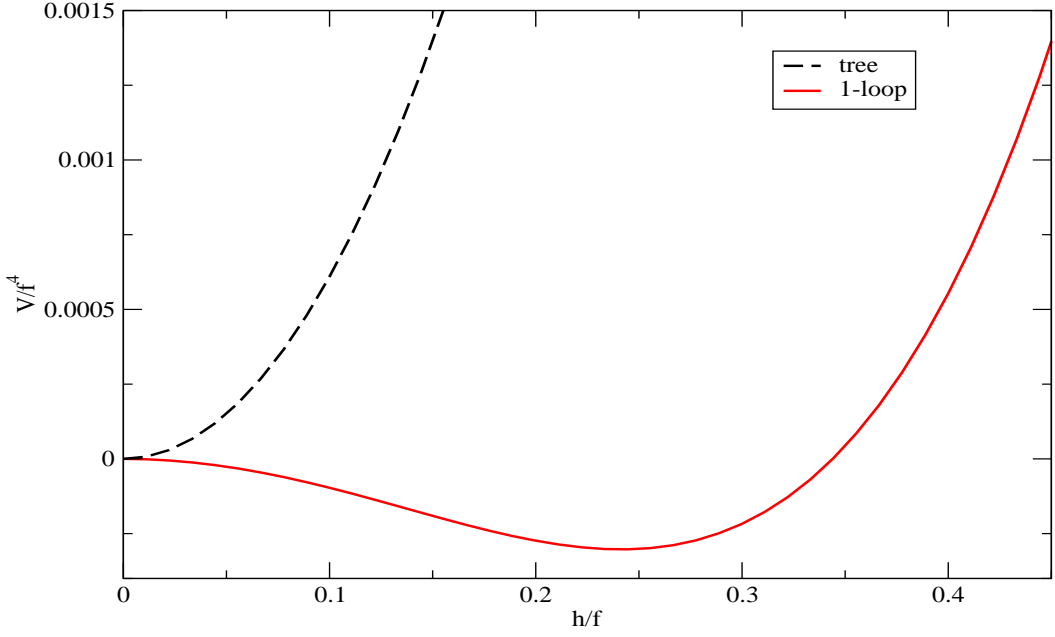,angle=-90,height=8cm,width=8cm,bbllx=4.cm,%
bblly=7.cm,bburx=20.cm,bbury=21.cm}}
\caption{\footnotesize
Close-up of fig.~\ref{fig:periodV} to reveal the electroweak minimum.}
\label{fig:ewmin}
\end{figure}

Therefore the breaking has to be triggered by one-loop radiative
corrections.  In order to compute the one-loop potential, both at $T=0$
and at finite $T$, we need the mass spectrum in an arbitrary Higgs
background $h$.  This is easy to compute and one finds that the masses
inherit the periodicity of $\langle \Sigma \rangle$ (we give the details
in Appendix~A). Then the one-loop potential is also a periodic
function of $h$, as shown by figure~\ref{fig:periodV}, where it can 
be compared with the tree level one.  In
particular, the height of the barriers between minima is somewhat reduced.
More importantly, the electroweak symmetry is now broken:  the
appearance of symmetry breaking minima is clearer in the close-up shown in
figure~\ref{fig:ewmin}. Analytically, this is understood as the result of
the negative contribution from the heavy top to the Higgs mass. To order
$h^2$ the $T=0$ one-loop potential reads
\bea
\delta_1 V &=&{h^2\over 64\pi^2}\left[-12\lambda_t^2 
M_{T}^2\left(\log{\Lambda^2\over M_{T}^2}+1\right)+{9\over 2} g^2 
M_{W'}^2\left(\log{\Lambda^2\over 
M_{W'}^2}+{1\over 3}\right)\right.\nonumber\\
&&\;\;\;\;\;\;\;\;\;\;\; + 6\left(\lambda_+ + 5 c 
{g'}^2-{\lambda_-^2\over 
\lambda_++3c{g'}^2}\right)M^2_{1}\left(\log{\Lambda^2\over M^2_{1}}
+1\right)\nonumber\\
&&\;\;\;\;\;\;\;\;\;\;\;
\left.+6\left(-\lambda_+ + {5\over 3} c {g'}^2+{\lambda_-^2\over 
\lambda_++3c{g'}^2}\right)M^2_{2}\left(\log{\Lambda^2\over M^2_{2}}
+1\right)\right]+...,
 \eea
where
\bea
M_{T}^2&\equiv &(\lambda_1^2+\lambda_2^2)f^2\ , \\
\label{MT}
M_{W'}^2&\equiv &{1\over 4}G_+^2f^2\ , \\
M^2_{1}&\equiv & 2(\lambda_++4c{g'}^2)f^2\ ,\\
M^2_{2}&\equiv & 2c{g'}^2f^2\ .
\label{M2}
\eea
We have chosen the following values for the parameters of the model:
$\lambda_1=\lambda_2=1$, $g_1=g_2=\sqrt{2}g$, $\lambda_+=0.85$ and
$\lambda_-=-0.0077$, so that $\langle h\rangle=246$ GeV for $f=1$
TeV.\footnote{The smallness of $\lambda_-$ gives a small vev to $t$,
which we then neglect to focus only on the $h$-direction. For the
behaviour of the potential along the $t$-direction see Appendix~B.} These
parameters also give adequate values for the masses of the non SM
particles in the model.

Before proceeding, it is convenient to say a word about our way of
computing the masses of scalar fluctuations ({\it i.e.} the degrees of
freedom in $\Sigma$ itself). In principle we could simply shift
$h^0\rightarrow h^0+h/\sqrt{2}$ in $\Sigma$ to compute $h$-dependent
scalar masses. To do this properly one should take into account that in
general, after this shifting, the scalar kinetic terms from (\ref{Lk}) are
not canonical. Therefore one should rescale the fields to get the kinetic
terms back to canonical form and this rescaling affects the ${\cal
O}(h^2)$ contributions to scalar masses. Instead of following this
standard procedure we find it convenient to use an alternative method
which simplifies the calculations and has appealing features when one is
concerned about the global structure of the potential, see below.

The idea is to treat the new background with $\langle H\rangle 
=h/\sqrt{2}$ as
a basis change (recall the discussion of the change from 
$\langle\Phi\rangle=I_5$ to $\langle\Phi\rangle=\Sigma_0$ in section 2). 
The $SU(5)$ 
transformation is now $U_h\equiv \exp (i\langle \Pi\rangle/f)$
with $\Sigma_h\equiv \langle \Sigma \rangle=U_h \Sigma_0 U_h^T$. To 
parametrize the scalar 
fluctuations around this 
background we again use the exponentials of broken generators, but taking 
into account the effect of the change of basis, which acts on generators 
as $X_a\rightarrow U_h X_a U_h^\dagger$. That is, instead of using 
$\exp(i\Pi/f)$ we use $U_h \exp(i\Pi/f) U_h^\dagger$,
and write for $\Sigma$:
\be
\Sigma = (U_h e^{i\Pi/f} U_h^\dagger)(U_h \Sigma_0 U_h^T)(U_h^* 
e^{i\Pi^T/f} U_h^T)
= e^{i\langle\Pi\rangle/f} e^{2i\Pi/f}e^{i\langle\Pi\rangle/f} 
\Sigma_0 \ . 
\label{good}
\ee
Alternatively, defining $\Pi_N = \pi_{a,N} X^{a,N}$ with  $X_{a,N} = \exp(i
\langle\Pi\rangle/f) X_a \exp(-i \langle\Pi\rangle^T/f)$ we have 
\be
\Sigma =  e^{i \Pi_N/f} \Sigma_h e^{-i \Pi^T_N/f} =   e^{2i \Pi_N/f} 
\Sigma_h\ .
\label{sigmaN}
\ee
[This parametrization allows the second
equality of eq.~(\ref{sigmaN}) to hold as the condition $X_{a,N} \Sigma_h
= \Sigma_h X_{a,N}^{T}$ is still valid.]
The prescription in eq.~(\ref{good}) [or (\ref{sigmaN})] is to be compared 
with the standard procedure
\be
\Sigma = e^{2i(\Pi+\langle\Pi\rangle)/f}\Sigma_0 \ .
\label{bad}
\ee
Some comments are in order. First, it is easy to check that with the 
prescription of eq. (\ref{good}) scalar fluctuations are automatically 
canonical. Second, we are free to choose this parametrization of scalar 
fluctuations: a general theorem \cite{coord} guarantees that this 
different 
parametrization does not change the physics. Finally, using 
eq.~(\ref{good}) 
has another advantage when looking at global properties of the 
scalar potential. As mentioned above, the mass spectrum is the same in 
all the 
periodic degenerate minima of the 
potential. For the scalar sector this is trivial to show with our 
parametrization and a bit more cumbersome using the conventional 
parametrization of eq.~(\ref{bad}) which, as explained, requires field 
redefinitions in order to recover canonical kinetic terms.
Some of these field redefinitions are in fact singular (because some 
kinetic terms go to zero in these minima). Although the singularities 
cancel out when computing scalar masses (because some masses also go to 
zero in these minima) they can make some of the scalar couplings blow up.
This can be interpreted as a failure of the parametrization (\ref{bad}) to 
cover the whole parameter space with a single coordinate patch. In this 
sense the parametrization (\ref{good}) is to be preferred when discussing 
global properties of the scalar potential.

\section{Effective Potential (${\bma T} {\bma \neq} {\bma 0}$) and discussion}

We calculate the finite temperature effective potential at one loop in
order to study the dynamics of the scalar fields at finite temperature
including the interactions with the thermal bath. To do this, we include
the contributions from all particles that receive a correction to their
mass from the vacuum expectation value of the nonlinear sigma model field
$\langle\Sigma\rangle \equiv \Sigma_h$, just as we did for the one-loop
potential at $T=0$. The one-loop thermal integrals for the contributions 
of bosonic and fermionic particles are standard, see e.g.~\cite{thermal}. 
For certain regions of $h$ some scalars might have negative masses squared, 
$m^2(h)<0$. For these we simply take the real part of the corresponding 
thermal integral\footnote{ In \cite{Wetterich} an alternative treatment of 
these integrals, with an infrared momentum cutoff $|m^2|$, is used. We 
have checked that both prescriptions are very close numerically.}.
\begin{figure}
\vspace{1.cm} \centerline{\vbox{
\psfig{figure=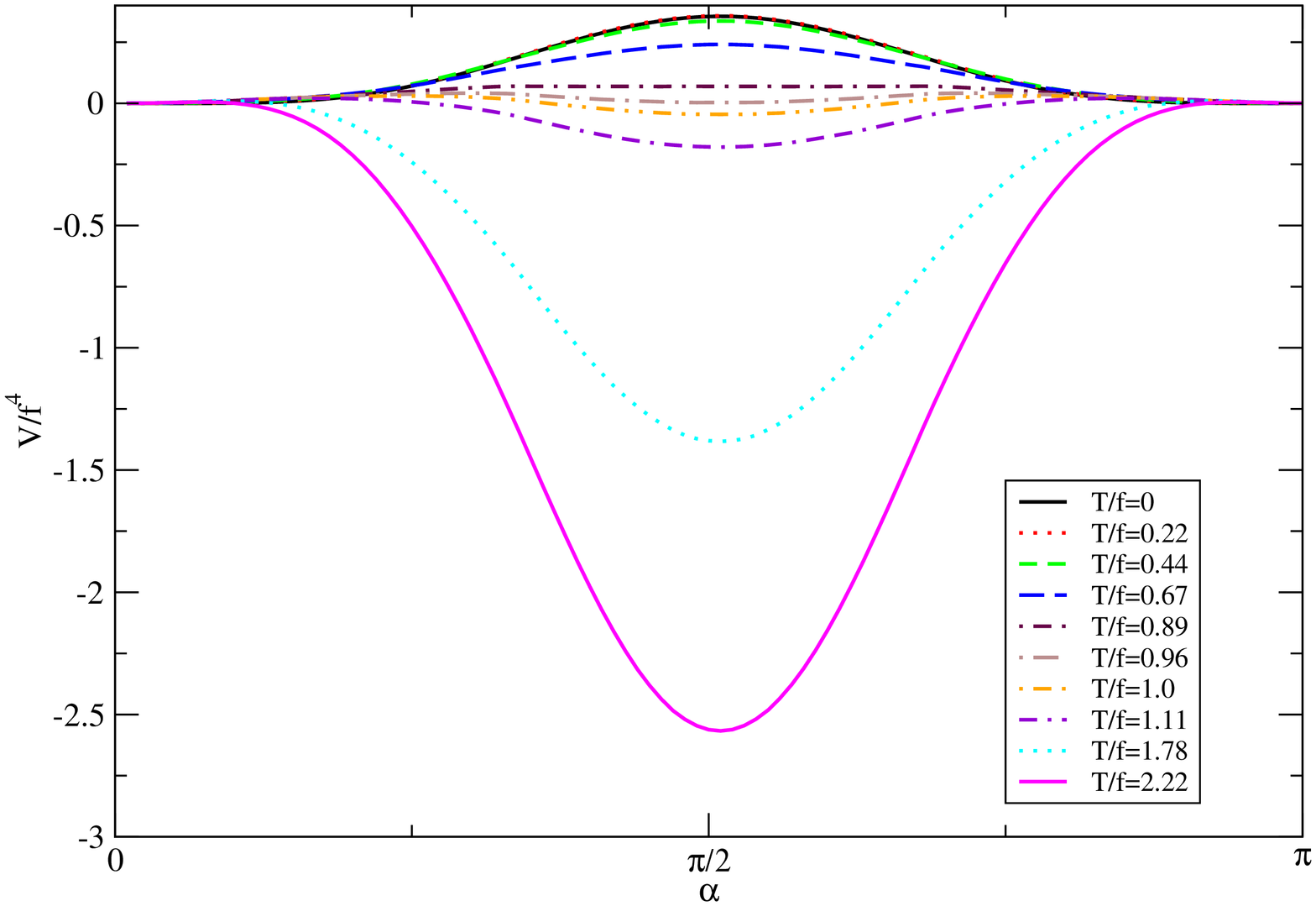,angle=-90,height=8cm,width=8cm,bbllx=7.cm,%
bblly=7.cm,bburx=23.cm,bbury=21.cm}
\psfig{figure=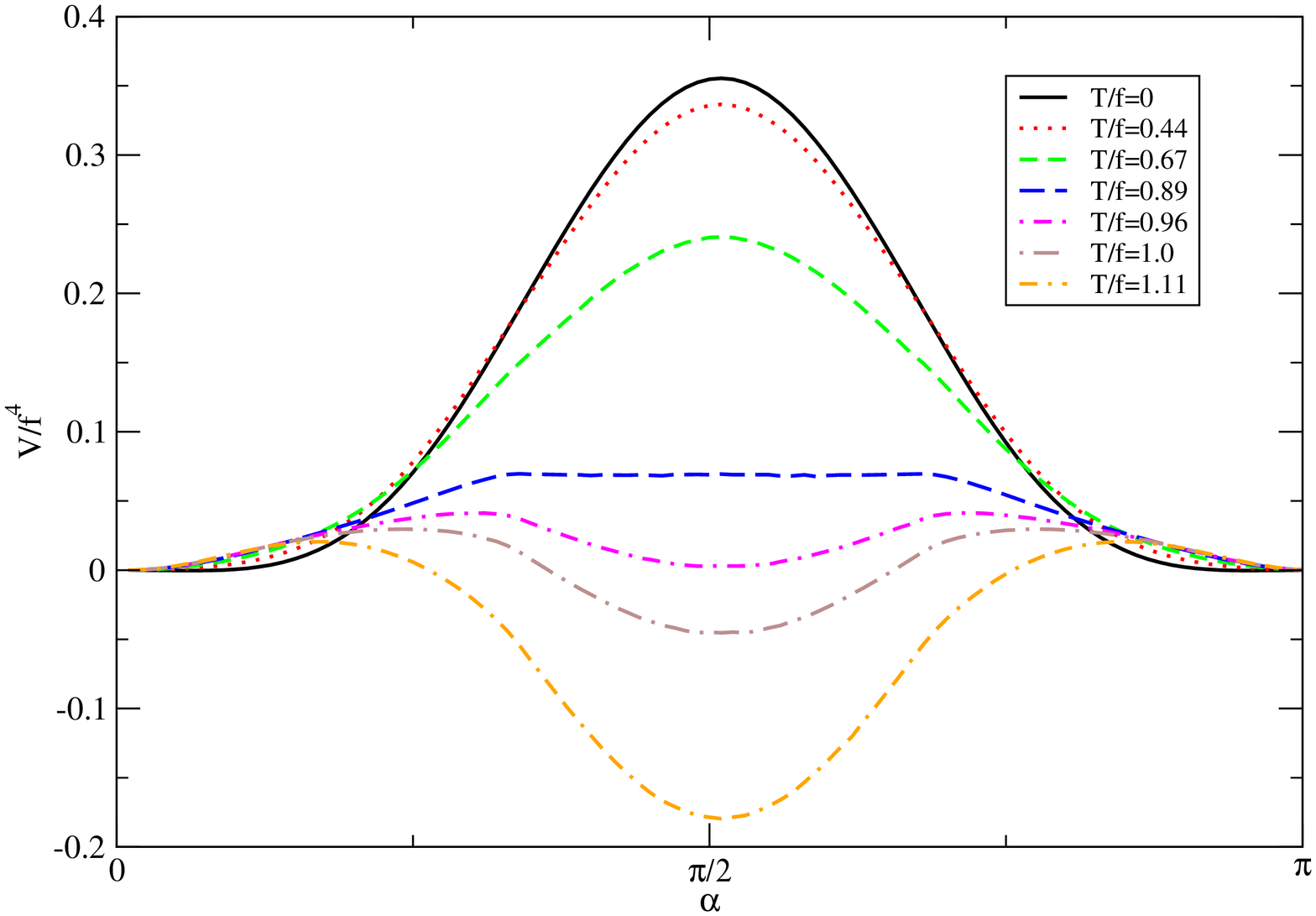,angle=-90,height=8cm,width=8cm,bbllx=4.cm,%
bblly=7.cm,bburx=20.cm,bbury=21.cm}
}}
\caption{\footnotesize
Evolution of the global structure of Higgs effective potential with 
temperature. The lower plot is a detail of the upper one, for a narrower 
range of temperatures around the critical temperature $T_1$.}
\label{fig:VT}
\end{figure}

Figure~\ref{fig:VT} shows the behaviour of the global structure of the
potential for increasing values of the temperature. We see that the
barrier between two minima decreases when $T$ is increased, eventually
turning into a local minimum (where the electroweak symmetry is broken)
that becomes the global minimum if $T$ is even higher. We can define a
critical temperature $T_1$ at which this local minimum is degenerate with
the minima at $h=0$. Numerically we obtain $T_1\simeq 0.96 f$ for our 
particular choice of parameters ($T_1\sim f$ will be generic). An analysis 
of the thermal behaviour of models with pseudo-Goldstone bosons was done 
in \cite{Holman}, in a different context. We find  that some of the 
generic features discussed in those papers resemble the behaviour  we 
encounter for Little Higgs models.

In order to understand analytically the particular behaviour of the 
potential shown in figure~\ref{fig:VT} it is enough to consider the case 
of $T\simgt 
f$. At such high $T$ (still below $\Lambda$) an expansion in 
powers of $m/T$ gives a good approximation to the potential. Writing also the 
one-loop $T=0$ part, one obtains in general
\bea
\delta_1 V&\simeq &{T^2\over 24}\left({\mathrm Tr} M_B^2 + {1\over 2}
{\mathrm Tr} M_F^2\right)-{T\over 12\pi} {\mathrm Tr} M_B^3
\nonumber\\
&+&{1\over 64\pi^2}\left[{\mathrm Tr} M_B^4\left(\log{c_B T^2\over 
\Lambda^2}+\kappa_B\right)-{\mathrm Tr} M_F^4\left(\log{c_F T^2\over
\Lambda^2}+\kappa_F\right)
\right]\ ,
\eea
where $B=\{S,V\}$ labels bosonic degrees of freedom (scalar and vector 
bosons) and $F$ fermionic ones. The constants $\kappa_{B,F}$ come from the 
one-loop $T=0$ potential and are simply $\kappa_S=\kappa_F=-3/2$ and 
$\kappa_V=-5/6$ while $c_{B,F}$ come instead from the finite temperature 
part and are $\log c_B\simeq 5.41$ and $\log c_F\simeq 2.64$. Keeping only 
the dominant $T^2$ term we get for our model (see appendix~A):
\be
\delta_T V=-{1\over 24}f^2 
T^2\left\{\left[12\lambda_++\lambda_1^2(64c'+6)
-{3\over 8}(g_1^2+g_2^2)\right]s_\alpha^4
+ {g'}^2\left[48c-{3\over 2}\right]s_\alpha^2\
\right\} \ .
\label{highTapp}
\ee
First we see that expanding around $h=0$, the thermal mass  is
proportional to $g'$, as expected on general grounds. For the range of
temperatures we are considering, this thermal effect is not strong enough 
to produce
a minimum in $h=0$, the one-loop $T=0$ corrections dominate and  indeed the
potential at $h=0$ has a maximum. Next we examine larger values of
$h$. In eq.~(\ref{highTapp}) the gauge contribution is subdominant, and
the scalar and fermionic terms provide a deep minimum for the potential at
the intermediate point $\alpha=\pi/2$.  This is precisely the high $T$
behaviour that we find in figure~\ref{fig:VT}.

The results indicating symmetry nonrestoration of the gauge 
electroweak symmetry directly lead us to pose the question of what could 
be the associated cosmology. As we have said already, we are restricted by our
analysis to temperatures beneath $\Lambda/\pi\simeq 4 f$.  
Suppose now that we follow the behaviour of the system starting from low
temperatures and heating up the thermal bath. As one can infer from the
lower plot of Fig.~\ref{fig:VT}, the electroweak gauge symmetry is
restored at some temperature $T_{\rm ew}$ of the order of the electroweak
scale. This means that $h=0$ (modulo $\pi f$)  becomes the ground state of
the system. This is completely analogous to what happens in the SM.
However, a further increase of the temperature 
results in a decrease of the height of the barrier between the periodic
minima at $h=0,\pi f,...$.  For even larger values of the temperature
beyond   $T_1\sim f$, the maximum at $h=\pi f/2 $ (modulo $\pi 
f$) turns into the global minimum and the barrier among the new equivalent 
vacua increases with temperature. 

The cosmology of the Little Higgs model will therefore depend strongly on
the maximum temperature that the Universe has attained after a period of
inflation \cite{lr}, which is necessary to explain the homogeneity and the
isotropy of our observed Universe.  Suppose that such a temperature is
smaller than $T_1$ and that the minima at $h =0,\pi f,...$ are always the
ground state until the electroweak phase transition.  
If so, the situation is completely analogous to the SM. Suppose, however,
that the highest temperature after inflation is indeed larger than
$T_1$. Under these circumstances, the Universe is characterized by
a high temperature phase during which the electroweak symmetry
is broken. This period will last till the temperature drops somewhat below 
$T_1$ at which point there will be a phase transition to the symmetric 
phase (followed later on by the usual SM electroweak phase transition 
back to the broken phase). This prediction is completely different from 
what is obtained in the SM. It would be very interesting to
analyze the role played by this non-standard phase of broken symmetry  
in the evolution of the early Universe, 
{\it e.g.} on the possible baryon asymmetry production \cite{rt}.

In view  of the small thermal mass that comes from (\ref{highTapp}) (which 
is even 
zero in most Little Higgs models), one might wonder if two-loop $T\neq 0$
effects can then become important\footnote{We thank Bob McElrath for 
pointing this out.}. In fact, at $T=0$, two-loop quadratically 
divergent corrections to the Higgs mass 
parameter go as $\delta m^2 \sim \Lambda^2/(16\pi^2)^2\sim f^2/(16\pi^2)$, 
the same order as one-loop non-divergent contributions. (We did not 
include  
them because the one-loop contributions will in principle dominate due to 
logarithmic enhancement). At $T\neq 0$ one would correspondingly expect 
corrections of 
order $\delta m_{eff}^2 \sim T^2/(16\pi^2)$, which may be important 
precisely when the one loop thermal mass is suppressed (or absent).
In fact, as the $T=0$ Higgs mass squared is itself of order 
$f^2/(16\pi^2)$, a 
two-loop thermal mass of that order can be relevant when $T\simgt f$. A 
complete two-loop 
calculation of the finite $T$ effective potential is beyond the scope of 
this paper, but it is of interest and we plan to undertake it in a future 
analysis. At this point we simply make two remarks. The first is that, for 
values of $h$ away from the origin, the one-loop $T^2$ contribution to the 
potential [eq.~(\ref{highTapp})] is negative and not suppressed. 
Therefore, the 
potential in that region will not change much after including two-loop 
corrections, which will be sub-dominant there. This gives us confidence on 
the inverse symmetry breaking behaviour we have found. The second remark 
is that we see no reason to expect that the two-loop contributions to the 
Higgs thermal mass will be positive. If they are, the details of the 
transition around $T_1$ will change (but not the existence of the 
transition itself) while if they are negative, our conclusions would be 
even stronger.

Let us close this section by a couple of comments. First, a complete study
of the scalar potential should also include the temperature evolution of
the triplet vev. For that purpose the spectrum in a more general
background with nonzero $h$ and $t$ is needed. This complicates
significantly the analysis, especially if one insists on keeping the $h$
and $t$ dependence to all orders (the potential is also periodic along the
$t$ direction). We nevertheless performed such analysis and some of our
results are presented in appendix~B. Schematically, the potential has an
egg-crate structure with different barrier heights for each direction.
Although the potential at finite temperature along the $t$ direction can
behave quite differently, depending on the choice of parameters, from what
we described for the $h$ direction we find it 
interesting at present to focus on the Higgs direction alone.

Secondly, the previous analysis has assumed that $f$ is a constant, 
independent of
temperature. However, we know that $f$ is in fact the vev of $\Sigma$
(along a particular direction), producing the breaking $SU(5)\rightarrow
SO(5)$. As such it is a dynamical variable and it is expected that at
sufficiently high $T$ one will get $f\rightarrow 0$, corresponding to a
critical temperature $T_*$ for the $SU(5)\rightarrow SO(5)$ transition. As
we do not know what is the physics beyond the cutoff scale $\Lambda$,
where the dynamics of the system is superseded by the UV completion of the
theory, we do not know the potential that produced the vev $f$ in the
first place. Therefore, it is difficult to be precise about $T_*$. However
we can make an estimate of the temperature behaviour of $f$ (in the spirit
of \cite{Wetterich} for the more complicated case of the chiral condensate
in QCD). We can just approximate the zero temperature potential for $f$ 
by a Mexican hat
potential
\be
V(f)=\kappa^2(f^2-\langle f\rangle^2)^2\ ,
\label{Vf}
\ee
where $\kappa$ is some unknown constant and $\langle f\rangle$ is assumed 
to be $\sim 1$ TeV (it corresponds to the $f$ used in the rest of the 
paper). We then add to the potential (\ref{Vf}) the finite temperature 
corrections coming from all the particles that have an $f$-dependent mass.
These are listed in eqs.~(\ref{MT})-(\ref{M2}). The value of 
$\kappa$ in eq.~(\ref{Vf}) is now crucial to the change of $f$ with 
increasing $T$. We argue that a natural choice is $\kappa\sim 2\pi$, 
because fluctuations around the minimum of the potential (\ref{Vf}) along 
the $f$ direction have mass $2\kappa \langle f\rangle$ and we are assuming 
that the only scalar fields below the cutoff scale $\Lambda=4\pi \langle 
f\rangle$ are those contained in $\Pi$. Therefore we should demand 
$2\kappa \langle f\rangle\simgt \Lambda$ which indeed translates into 
$\kappa\simgt 2\pi$. Choosing then that value of $\kappa$ for the 
numerical evaluation of the effective potential for $f$, we find that 
$f$ changes very little with $T$. For the extreme value $T=4\langle 
f\rangle$ we obtain $f(T)\simeq 0.9 f$, just a $10\%$ decrease.

\section{Conclusions}

We have shown, on very general grounds, that the behaviour of Little Higgs 
models at finite temperature is considerably richer than in the Standard Model.
In particular we have studied the effective  potential at finite  
temperature which, as the Higgs $h$ in these models is a pseudo-Goldstone 
boson, is a periodic function of $h$. 

Although the electroweak phase transition is expected to occur just like in
the Standard Model, at higher temperatures, when the new states introduced
in Little Higgs models at a scale $f\sim 1$ TeV are thermally produced in
the plasma, the history of the early Universe changes dramatically. At
 some temperature $T_1\sim f$ a new minimum where the electroweak
symmetry is broken becomes the global minimum of the potential
and the gauge electroweak symmetry becomes more and more broken as the
 temperature continues to increase. Being a perturbative statement, it would
be interesting to see if such a behaviour persists when non-perturbative
corrections are accounted for, or what is the fate of this broken minimum 
in the context of the UV completion of the theory. 
At any rate, this rich structure at high temperatures
 might have cosmological implications which are worth  studying.

As a spin-off, we have introduced a new parametrization of scalar 
fluctuations around displaced vacua in nonlinear sigma models which has 
very appealing features and might be of interest for phenomenological 
studies.

\section*{Acknowledgments}

We acknowledge very useful discussions with (and help from) Alberto Casas, 
Mikko Laine, Bob McElrath, Jes\'us Moreno and Ann Nelson. This work was 
supported in 
part by Colciencias 
under contract no.  1233-05-13691. J.R.E. thanks CERN and Univ. of Padova 
for hospitality during several stages of this work.  M.L.  thanks the 
I.E.M. of CSIC-Madrid and LPT-Universit\'e de Paris XI-Orsay for 
hospitality
during the completion of this work. A.R. thanks CERN for hospitality.

\section*{A. Spectrum of the Little Higgs model}
\setcounter{equation}{0}
\renewcommand{\theequation}{A.\arabic{equation}}

In this appendix we present the mass matrices in a Higgs background for
the Little Higgs model \cite{Peskin} studied in this paper. These masses
are needed in the calculation of the one-loop Higgs potential, both at
zero and finite temperature. We keep the exact dependence on the Higgs
background $\langle h^0\rangle = h/\sqrt{2}$ so that we can study the
global structure of the potential. We also present the mass eigenvalues in
an expansion up to $h^2$. The scalar sector contains a complex $SU(2)$
triplet, a real triplet, the Higgs doublet and a singlet, a total of 14
degrees of freedom that are distributed in one doubly charged scalar,
three charged fields and 6 neutral and real fields (4 CP odd and 2 CP
even). The gauge boson sector contains a heavy $W'$ and a heavy $Z'$
besides the SM gauge bosons. The fermion sector of relevance for our
calculation contains a heavy top in addition to the SM top quark. We
follow the notation introduced in section~2. We remind the reader here that
$\lambda_+\equiv c\,G_+^2+16c'\lambda_1^2$ and $\lambda_-\equiv
c\,G_-^2-16c'\lambda_1^2$, with $G_+^2\equiv g_1^2 + g_2^2$, $G_-^2\equiv
g_1^2 - g_2^2$. We also use $\alpha \equiv h/(\sqrt 2 f)$, where $h$ is
normalized as a real field. (At $T=0$ one has $\langle h\rangle=246$ GeV.)
The mass matrices presented below correspond to canonically normalized 
fields (see section~3 for a discussion on this point for the scalar 
sector). The mass eigenvalues of all scalar fields are periodic under 
$\alpha \rightarrow \alpha + \pi$.

\subsection*{Doubly charged scalar}
The field $\phi^{++}$ has a  mass
\be
M_{\phi^{++}}^2=2f^2(\lambda_+ + 4 c {g'}^2 - 16 c'\lambda_1^2 s^4_\alpha )
=2(\lambda_+ + 4 c {g'}^2)f^2+ {\cal O}(h^4/f^2)\ .
\ee
At order $h^2$ this state 
does not contribute to the trace of the mass squared operator.

\subsection*{Charged scalars}

In the basis $\{\xi^+,\phi^+,h^+\}$, the mass matrix
$M_+^2$ for the charged scalars $\varphi_i^+$ is
\bea
M^2_+&=&f^2\left[
\begin{array}{ccc}
-2\lambda_+c_\alpha(2+c_\alpha)s^4_{\alpha/2} & 
{1\over 2}\lambda_+c_\alpha^2 s^2_{\alpha}
& -{i\over \sqrt{2}}\lambda_- s_{2\alpha}s^2_{\alpha/2}\\
&&\\
{1\over 2}\lambda_+c_\alpha^2 s^2_{\alpha}
& 2\lambda_+c_\alpha(2-c_\alpha)c^4_{\alpha/2} & 
-{i\over \sqrt{2}}\lambda_- s_{2\alpha}c^2_{\alpha/2}\\
&&\\
{i\over \sqrt{2}}\lambda_- s_{2\alpha}s^2_{\alpha/2} &
{i\over \sqrt{2}}\lambda_- s_{2\alpha}c^2_{\alpha/2} &
\lambda_+ s^2_{\alpha} c^2_{\alpha}
\end{array}
\right]\nonumber\\
&&\nonumber\\
&+& c{g'}^2 f^2\left[
\begin{array}{ccc}
2(1-3c_\alpha)s^2_{\alpha/2} & s^2_{\alpha} & 0 \\
s^2_{\alpha} & 2(1+3c_\alpha)c^2_{\alpha/2} & 0 \\
0 & 0 & 2 c^2_{\alpha}
\end{array}
\right]\ .
\eea
For the trace we obtain
\be
2{\mathrm Tr}
M^2_+= 4\lambda_+ f^2(1-s^4_\alpha)+4c{g'}^2 f^2(5-4s^2_\alpha)\ ,
\ee
and an expansion in powers of $h$ gives
\be
2{\mathrm Tr}
M^2_+= {\mathrm Constant} -8c{g'}^2 h^2 + {\cal O}(h^4/f^2)\ ,
\ee
with the only contribution to order $h^2$ being that of the $U(1)_Y$ 
sector.
A similar expansion for the mass eigenvalues gives
\bea
m^2_{\varphi_1^+}&=&-{1\over 2} c{g'}^2h^2+...\\
m^2_{\varphi_2^+}&=&2(\lambda_++4 c{g'}^2)f^2 -{1\over 2} \left[\lambda_++5 
c{g'}^2-{\lambda_-^2\over \lambda_++3 c{g'}^2}\right]h^2+...\\
m^2_{\varphi_3^+}&=&2c{g'}^2f^2+{1\over 2}\left[ 
\lambda_+-2c{g'}^2-{\lambda_-^2\over \lambda_++3 
c{g'}^2}\right]h^2+... 
\eea
For $h=0$, $\varphi_1^+=\xi^+$ is the charged Goldstone boson associated to the
gauge symmetry breaking $SU(2)_1\times SU(2)_2\rightarrow SU(2)_{SM}$ and
we get $m^2_{\varphi_1^+}=0$.

\subsection*{Neutral scalars}

For the neutral scalar fields $\varphi_i^0$ we use the basis
$\{\zeta^0,\xi^0,\phi^{0r},\phi^{0i},h^{0r},h^{0i}\}$, where we have
decomposed the complex fields $\phi^0$ and $h^0$ as $h^0=(h^{0r}+i
h^{0i})/\sqrt{2}$ and $\phi^0=i(\phi^{0r}+i\phi^{0i})/\sqrt{2}$.  The mass
matrix for neutral scalar fields, $M_{0}^2$, breaks up in two blocks:  
one for the pseudoscalars $\{\zeta^0,\xi^0,\phi^{0i},h^{0i}\}$ and the
other for the scalars $\{\phi^{0r},h^{0r}\}$. The $4\times 4$ block is
\bea
M^2_{4\times 4}&=&-{1\over 4}f^2\left[
\begin{array}{cccc}
 5\lambda_+s_\alpha^4 & 
 \sqrt{5}\lambda_+s_\alpha^4 & 
\sqrt{10}\lambda_+(2-s_{\alpha}^2)s^2_{\alpha} & 0\\
\sqrt{5}\lambda_+s_\alpha^4 &
\lambda_+s_\alpha^4 &
\sqrt{2}\lambda_+(2-s_{\alpha}^2)s^2_{\alpha} & 0\\
\sqrt{10}\lambda_+(2-s_{\alpha}^2)s^2_{\alpha} & 
\sqrt{2}\lambda_+(2-s_{\alpha}^2)s^2_{\alpha} &
-2 \lambda_+(4c_\alpha^2- s^4_{\alpha})
& 4\lambda_-c_\alpha s_{2\alpha}\\
0 & 0 & 4\lambda_-c_\alpha s_{2\alpha} & -\lambda_+ s^2_{2\alpha}
\end{array}
\right]\nonumber\\
&&\nonumber\\
&-& c{g'}^2 f^2\left[
\begin{array}{cccc}
5s_\alpha^2 &
\sqrt{5}s_\alpha^2 & 
\sqrt{10}s^2_{\alpha} & 0\\
\sqrt{5}s_\alpha^2 &
s_\alpha^2 &
\sqrt{2}s^2_{\alpha} & 0\\
\sqrt{10}s^2_{\alpha} &    \sqrt{2}s^2_{\alpha}& -2(4-5s^2_{\alpha}) & 0\\
0& 0 & 0  &  -2c^2_{\alpha}
\end{array}
\right]\ ,
\eea
which has a zero eigenvalue, corresponding to the eigenvector 
$(1,-\sqrt{5},0,0)/\sqrt{6}$ [the neutral Goldstone boson associated to the 
breaking of an ungauged $U(1)$].

The $2\times 2$ block is:
\be
M_{2\times 2}^2=f^2\left[
\begin{array}{cc}
\lambda_+(2c_\alpha^2-s_\alpha^4) & \lambda_-(3s_\alpha^2-2)s_\alpha\\
\lambda_-(3s_\alpha^2-2)s_\alpha & \lambda_+ (3-4s_\alpha^2)s_\alpha^2
\end{array}
\right]+ 2c{g'}^2f^2\left[
\begin{array}{cc}
4-5s_\alpha^2 & 0 \\
0 & c_{2\alpha}\end{array}
\right]
\ .
\ee
From these matrices we obtain
\be
{\mathrm Tr} M^2= 4\lambda_+f^2 (1-2s^4_\alpha)
+ 4 c{g'}^2 f^2(5-8 s^2_\alpha) \ .
\ee
An expansion in powers of $h$ gives
\be
{\mathrm Tr} M^2={\mathrm Constant}-16c{g'}^2 h^2 +{\cal O}(h^4/f^2)\ ,
\ee
with the only contribution of order $h^2$ being that of the $U(1)_Y$ 
sector.
The expansion of the mass eigenvalues gives, for the $4\times 4$ block:
\bea
m^2_{\varphi_1^0}&=&0\\
m^2_{\varphi_2^0}&=&-3 c{g'}^2h^2+...\\
m^2_{\varphi_3^0}&=&2(\lambda_++4 c{g'}^2)f^2 - \left[\lambda_++5 
c{g'}^2-{\lambda_-^2\over
\lambda_++3c{g'}^2}\right]    h^2  +...\\
m^2_{\varphi_4^0}&=&2c{g'}^2f^2+{1\over 2} \left[\lambda_+-2c{g'}^2-
{2\lambda_-^2\over
\lambda_++3c{g'}^2}\right]h^2+...
\eea
while for the $2\times 2$ block we get
\bea
m^2_{\varphi_5^0}&=&2(\lambda_++4 
c{g'}^2)f^2-\left[\lambda_++5c{g'}^2-{\lambda_-^2\over
\lambda_++3c{g'}^2}\right]h^2+...
\\
m^2_{\varphi_6^0}&=&2c{g'}^2f^2+{1\over 2}
\left[3\lambda_+-4c{g'}^2-{2\lambda_-^2\over 
\lambda_++3c{g'}^2}\right]h^2+...
\eea
For $h=0$ we have two zero mass eigenvalues: the Goldstones $\xi^0$ and 
$\zeta^0$.

\subsection*{Gauge Bosons}

The mass matrix for the charged gauge bosons ($W^\pm,{W'}^\pm$) 
in the interaction basis is given by 
\be
M_W^2={f^2\over 4}\left[
\begin{array}{cc}
g_1^2 & -g_1 g_2 c^2_\alpha \\
-g_1 g_2 c^2_\alpha  & g_2^2
\end{array}
\right]\ ,
\ee
which has periodic mass eigenvalues.

There are three neutral gauge bosons: $Z^0$, ${Z'}^0$ and the photon.
Their $3\times 3$ mass matrix in the interaction basis is
\be
M_{Z,A}^2={f^2\over 8}\left[
\begin{array}{ccc}
g_1^2 (2+s^4_\alpha) & -g_1 g_2(1+c^4_\alpha) & 2 g_1 g' s^2_\alpha\\
-g_1 g_2(1+c^4_\alpha)  & g_2^2 (2+s^4_\alpha) & 2 g_2 g' s^2_\alpha\\
2 g_1 g' s^2_\alpha & 2 g_2 g' s^2_\alpha & 4 {g'}^2 s^2_\alpha
\end{array}
\right]\ ,
\ee
again with periodic eigenvalues. It can be easily checked that $M_{Z,A}^2$ 
annihilates the vector $(1/g_1,1/g_2,-1/g')$, which corresponds to the 
photon. For the trace of $M^2$ we get
\be
{\mathrm Tr} M_V^2 =  2 {\mathrm Tr} M_W^2 + {\mathrm Tr} M_{Z,A}^2= {1\over 
8}G_+^2f^2(6+s_\alpha^4)+{1\over 
2} {g'}^2f^2s_\alpha^2\ ,
\ee
and expanding in powers of $h$,
\be
{\mathrm Tr} M_V^2 = {\mathrm Constant} + {1\over 4} {g'}^2h^2 + {\cal 
O}(h^4/f^2)\ .
\ee


The expansion of the mass eigenvalues is
\bea
m^2_{W^\pm}&=&{1\over 4} g^2h^2+...\\
m^2_{{W'}^\pm}&=&{1\over 4} G_+^2f^2-{1\over 4} g^2h^2...\\
m^2_{Z^0}&=&{1\over 4} (g^2+{g'}^2)h^2+...\\
m^2_{{Z'}^0}&=&{1\over 4} G_+^2f^2-{1\over 4} g^2h^2...
\eea
where we have used $g^2\equiv g_1^2g_2^2/(g_1^2+g_2^2)$.

\subsection*{Fermions}

For the top quark and its heavy partner we have
\be
M_{f}M_{f}^{\dagger} = f^2 \left( 
\begin{array}{cc}
\lambda_2^2 & \lambda_1 \lambda_2 c^2_\alpha \\
 \lambda_1 \lambda_2 c^2_\alpha & \lambda_1^2 (1 - s^4_\alpha)
\end{array}
\right)\ ,
\ee
which has periodic eigenvalues. The fermionic trace is
\be
{\mathrm Tr} (M_{f}M_{f}^{\dagger}) = f^2\left[\lambda_2^2+\lambda_1^2 (1-
s_{\alpha}^4)\right]\ ,
\ee
which does not contribute to order $h^2$.
The expansion of the two  mass eigenvalues can be written as
\bea
m^2_{t}&=&{1\over 2} \lambda_t^2h^2+...\\
m^2_{T}&=&(\lambda_1^2+\lambda_2^2)f^2-{1\over 2} \lambda_t^2h^2...
\eea
where the top Yukawa is given by $\lambda_t^2\equiv 
2\lambda_1^2\lambda_2^2/(\lambda_1^2+\lambda_2^2)$.

\section*{B. The triplet direction}
\setcounter{equation}{0}
\renewcommand{\theequation}{B.\arabic{equation}}

As mentioned in the main text, in this model a nonzero vacuum expectation
value of $\langle h^0 \rangle =h/\sqrt{2}$ will induce a vev for the
triplet field $\langle \phi_{0}^{r} \rangle=t/\sqrt{2}$. 
The tree-level potential to all orders in $h$ and $t$ is 
given by,
\bea
V_{0}&=& \frac{\lambda_+ f^4}{(t^2+h^2)^2}\left[ 3(t^2+h^2)^2 + 
\frac{1}{2} t^2(t^2+h^2) s^2_{2\beta} + \frac{1}{2}h^4 
s^4_\beta\right] \nonumber \\
&-&
\frac{ \lambda_- f^4}{(t^2+h^2)^2} t h^2 (t^2+h^2)^{1/2} s^2_\beta 
s_{2\beta} + \frac{2c f^4 g'^2}{(t^2+h^2)} \left[ t^2 s^2_{2\beta}  + 
h^2 
s^2_\beta\right]\ ,
\label{vtree}
\eea
where $\beta =(t^2+h^2)^{1/2}/(\sqrt{2} f)$. We see that the potential is 
 periodic along the $t$-direction as well.

A complete analysis of the one-loop corrections to the potential
(\ref{vtree}), both at $T=0$ and $T\neq 0$ requires the calculation of the
mass matrices of all particles in the presence of the background fields
$h,t$, which is not an easy task, although we did perform it.
The generic expressions one obtains for these mass matrices are too long 
and complicated to be given here. 

In this paper we limit ourselves to give an example of the kind of
dynamics one might obtain along the triplet direction. The interest of
this is not limited by the fact that the phenomenological complications
associated to the presence of a triplet vacuum expectation value can be
neatly solved by imposing a $T$-parity on the model \cite{ChengLow}.  As
we will explain next, even if one does not have a tadpole for $t$ the
triplet direction might play a role in the thermal evolution of the
Universe.

The argument that leads one to expect a peculiar behaviour of 
the effective potential at finite temperature along the Higgs doublet direction 
(namely the fact that quadratically divergent corrections to the Higgs 
mass vanish) does not hold for the triplet direction. Explicitly, an expansion of 
 the $t$-dependent masses for the gauge bosons
along the triplet direction gives
\bea
m^2_{W^\pm}&=&{1\over 2} g^2t^2+...\\
m^2_{{W'}^\pm}&=&{1\over 4} G_+^2f^2-{1\over 2} g^2t^2...\\
m^2_{Z^0}&=&(g^2+{g'}^2)t^2+...\\
m^2_{{Z'}^0}&=&{1\over 4} G_+^2f^2+{1\over 4} {G_-^4\over G_+^2}t^2...
\eea
so that the trace is
\be
3{\mathrm Tr}M_V^2={\mathrm Constant} + {3\over 4}(G_+^2+4{g'}^2)t^2+...
\ee
For fermions one gets
\be
12{\mathrm Tr}M_F^2={\mathrm Constant} - 24 \lambda_1^2 t^2+...,
\label{trfer}
\ee
while for scalars
\be
{\mathrm Tr}M_S^2={\mathrm Constant} - {4\over 5}(318 c{g'}^2+1488 
c'\lambda_1^2+
49 \lambda_+)t^2+...
\label{trscal}
\ee

These results determine that the thermal potential along the triplet direction 
has a $T^2$ contribution 
\bea
\delta V(T)&=&-{1\over 2}\kappa_t^2 t^2T^2\nonumber\\
&=&{1\over 48}
\left[{3\over 2}(G_+^2+4{g'}^2)-{8\over 5}(318 c{g'}^2
+1488 c'\lambda_1^2+49 \lambda_+)-24\lambda_1^2\right]t^2T^2\ .
\label{VTT}
\eea
The negative contributions to the traces in (\ref{trfer}) and 
(\ref{trscal}) dominate in the potential (\ref{VTT}), making 
$\kappa_t^2>0$. This unusual behaviour implies that for temperatures 
above a critical value
\be
T_t^c=M_t/\kappa_t\ ,
\ee
where $M_t^2=2(\lambda_+ + 4 c {g'}^2)f^2$ is the triplet mass squared,
the electroweak symmetry gets broken along the triplet direction. That is, 
even if a $T$-parity forbids a tadpole $h^2t$ in the potential, and even 
if the mass of the triplet is large (of order $f$), thermal corrections
will trigger electroweak symmetry breaking along the triplet direction, 
for some $T_t^c\sim f$. Of course one can increase the value of $T_t^c$ by 
appropriately choosing the parameters of the model and in this spirit we 
have not discussed this transition in the main text, and have focused 
instead in the Higgs direction. Nevertheless, this makes the 
possible structure of phase transitions in these models even richer.

\end{document}